\documentclass[twocolumn,tightenlines,prl,floatfix,nofootinbib]{revtex4}

 \usepackage[dvips,final]{graphicx}
  \usepackage{amssymb}
   \usepackage{amsmath}
    \usepackage{amsfonts}
     \usepackage{epsfig}
      \usepackage{bm}% bold math

\begin{document}

\title{Exclusive $b \bar b$ pair production
and irreducible background\\ to the exclusive Higgs boson
production}

\author{Rafal Maciu{\l}a$^1$}
\author{Roman Pasechnik$^2$}
\author{Antoni Szczurek$^{1,3}$}
 \affiliation{$^1$Institute of Nuclear Physics PAN, PL-31-342 Cracow, Poland}
 \affiliation{$^2$Uppsala University, Box 516, SE-751 20 Uppsala, Sweden}
 \affiliation{$^3$University of Rzesz\'ow, PL-35-959 Rzesz\'ow, Poland}

\begin{abstract}
We calculated the cross section for the exclusive double
diffractive $b\bar b$ production in $pp\to p(b\bar b)p$
reaction at LHC.
Large cross sections are obtained (3 -- 10 nb).
This process constitutes the irreducible background to the exclusive
Higgs production and is of particular importance in the upcoming
Higgs boson searches. The distribution in invariant mass of the $b
\bar b$ pair is calculated and compared with the corresponding
contribution from the Higgs decay. The contribution from the
exclusive production of $Z^0$ and its decay
as well as the contribution from the $\gamma \gamma \to b \bar b$
subprocess are also presented for the first time.
The influence of cuts on the signal-to-background ratio is discussed.
\end{abstract}

\maketitle

The identification of the Higgs boson produced in high-energy
proton-proton collisions is one of the main goals of the LHC
program. The dominant mechanisms of its production are the
gluon-gluon and WW fusions. The Higgs boson will be searched for in
different decay channels: $b \bar b$, $\gamma \gamma$, $\tau^+
\tau^-$, $W^+ W^-$, etc., in typical inclusive measurements, i.e.
when only the selected decay channel is studied, and the Higgs boson
is produced in association with many other particles.
Depending on the Higgs mass, the different channels seem more
favorable, and each of the decay channels has its own difficulties
for the experimental identification.

\begin{figure}
\includegraphics[width=8.0cm]{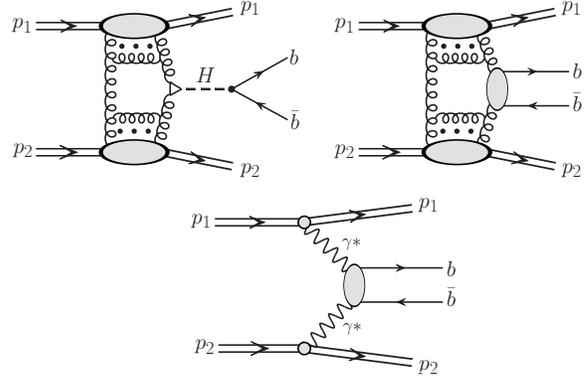}
\caption{The diagrams for the $b \bar b$ pair production
proceeding through the Higgs boson (top-left),
the exclusive diffractive $b\bar b$
production (top-right) and photon-photon fusion (bottom).}
\label{fig:diagrams}
\end{figure}

An alternative solution was suggested already some time ago. It was
proposed to identify the Higgs boson in exclusive process $pp\to
pHp$ in the central rapidity region \cite{Nachtmann,BL,Cudell}.
Modern estimates of the corresponding cross section in the
$k_t$-factorization approach with the help of unintegrated gluon
distribution functions (UGDFs) have been presented in
Ref.~\cite{KMR_Higgs}. It was argued that in the forward
scattering limit the background in the $b\bar b$ channel should be
small due to the so-called ``$J_z$ = 0 selection rule'' and a
suppression for massless quarks in this channel (see, for example,
Ref.~\cite{KMR_bbar_suppression} and references therein). In
general, there is also contribution in the $J_z$ = 2 channel and the
$b$-quarks are rather heavy in comparison with the typical soft
hadronic scale. Therefore, a realistic estimate of the
background requires a real calculation for the genuine four-body
reaction $pp\to p(b\bar b)p$. The diffractive mechanism
of the exclusive $b \bar b$ production is shown in
Fig.~\ref{fig:diagrams}. For comparison, we also show
the mechanism with intermediate Higgs boson, which is
a signal in our analysis and a reaction with the
$\gamma \gamma \to b \bar b$ subprocess.
%-------------------------------------------------------
%-------------------------------------------------------

The amplitude for the exclusive process $pp\to p(q\bar q)p$ can be
written as \cite{KMR_Higgs}
\begin{eqnarray}
{\cal M}_{\lambda_b\lambda_{\bar{b}}}=
\frac{s}{2}\cdot\frac{\pi^2\delta_{c_1c_2}}{N_c^2-1}\, \Im\int d^2
q_{0,t} \; V_{\lambda_b\lambda_{\bar{b}}}^{c_1c_2}(q_1, q_2, k_1, k_2) \nonumber \\
\times\frac{f^{\mathrm{off}}_{g,1}(x_1,x_1',q_{0,t}^2,
q_{1,t}^2,t_1)f^{\mathrm{off}}_{g,2}(x_2,x_2',q_{0,t}^2,q_{2,t}^2,t_2)}
{q_{0,t}^2\,q_{1,t}^2\, q_{2,t}^2}, \label{amplitude}
\end{eqnarray}
where $\lambda_q,\,\lambda_{\bar{q}}$ are helicities of heavy $q$
and $\bar{q}$, respectively, $t_{1,2}$ are the momentum transfers
along each proton line, $q_{1,t}, q_{2,t}, x_{1,2}$ and
$q_{0,t},\,x_1'\sim x_2'\ll x_{1,2}$ are the transverse momenta and
the longitudinal momentum fractions for active and screening gluons,
respectively. Above $f^{\mathrm{off}}_{g,1/2}$ are the off-diagonal
UGDFs of both nucleons. In the calculations presented here we take
$\mu^2 = M_{b \bar b}^2$, where $M_{b \bar b}$ is the invariant mass
of the $b \bar b$ system. This is consistent with using $\mu^2 =
M_H^2 = M_{b \bar b}^2$ for exclusive Higgs boson production, which
is preferred from the theoretical point of view
\cite{Forshaw_recent}. The hard $g^*g^* \to b \bar b$ subprocess
amplitude $V_{\lambda_q\lambda_{\bar{q}}}$ consists of the Higgs
decay signal $g^*g^*\to H\to b\bar b$ and background contributions,
where the major ones come from the direct $g^*g^*\to b\bar b$
process, from the photon-photon fusion and from the exclusive
production of $Z^0$ and its subsequent decay $Z^0 \to b \bar b$.
Note, that due to integration over ${\bf q}_{0,t}$ in the
diffractive amplitude (\ref{amplitude}), only symmetric part of the
hard subprocess amplitude $V_{\lambda_q\lambda_{\bar q}}({\bf
q}_{0,t})= V_{\lambda_q\lambda_{\bar q}}(-{\bf q}_{0,t})$
contributes to the diffractive cross section.

The off-diagonal UGDFs are written as \cite{KMR-ugdf}
 \begin{equation}
 f^{\mathrm{off}}_g(x',x_{1,2},q_{1,2t}^2,q_{0,t}^2,\mu_F^2)\simeq
 R_g\,f_g(x_{1,2},q_{1,2t}^2,\mu_F^2),
 \label{rg}
 \end{equation}
where $R_g\simeq 1.2$ which accounts for the single $\log Q^2$
skewed effect \cite{Shuvaev:1999ce}. In the considered kinematics
the diagonal unintegrated densities can be written in terms of the
conventional (integrated) densities $xg(x,q_t^2)$ as~\cite{KMR-ugdf}
\begin{equation}\label{ugdfkmr}
f_g(x,q_t^2,\mu^2)=\frac{\partial}{\partial\ln q_t^2}
\big[xg(x,q_t^2)\sqrt{T_g(q_t^2,\mu^2)}\big] \; ,
\end{equation}
where $T_g$ is the conventional Sudakov survival factor which
suppresses real emissions from the active gluon during the
evolution, so the rapidity gaps survive. The gluon $q_t$'s typical
for the central exclusive Higgs production at LHC are of the order
of few GeV \cite{KMR_Higgs}.

Additionally, following to Ref.~\cite{Forshaw_recent} we use the
factorization scale $\mu_F=M_{b{\bar b},t}$ given by the transverse
mass of the $b{\bar b}$ pair $M_{b{\bar b},t}$ as compared to the
KMR convention \cite{KMR_Higgs} $\mu_F^{\text{KMR}}=M_{b{\bar
b},t}/2$. We will discuss uncertainties due to sensitivity to the
factorization scale $\mu_F$ choice and conventional gluon densities
$xg(x,q_t^2)$ at small $x$ and $q_t$ below when presenting numerical
results.

\begin{figure*}[tbh]
 \centerline{\includegraphics[width=0.8\textwidth]{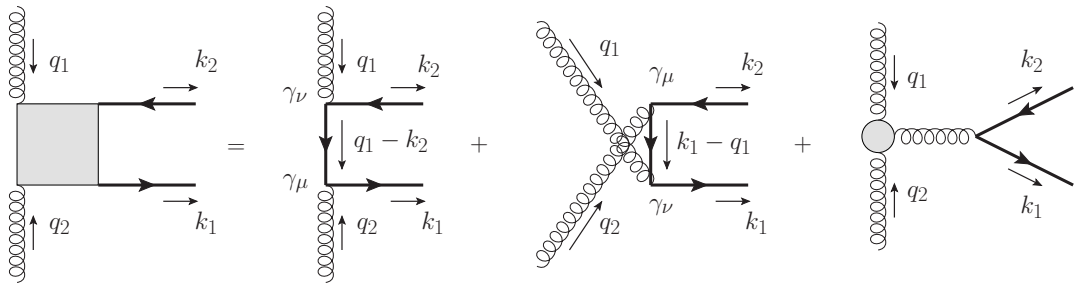}}
 \caption{Effective vertex in the QMRK approach \cite{FL96}. Last
   diagram with effective 3-gluon vertex drops out in projection to
the color singlet final state.}
 \label{flvertex}
\end{figure*}

In the framework of the $k_t$-factorization approach
\cite{ktfac} the hard subprocess $g^*g^*\to q\bar q$ gauge invariant
amplitude reads
\begin{eqnarray}\nonumber
&&V_{\lambda_q\lambda_{\bar{q}}}^{c_1c_2}(q_1,q_2)\equiv
n^+_{\mu}n^-_{\nu}V_{\lambda_q\lambda_{\bar{q}}}^{c_1c_2,\,\mu\nu}(q_1,q_2),\quad
n_{\mu}^{\mp}=\frac{p_{1,2}^{\mu}}{E_{p,cms}},\\
&&V_{\lambda_q\lambda_{\bar{q}}}^{c_1c_2,\,\mu\nu}(q_1,q_2)=-g_s^2\sum_{i,k}\left\langle
3i,\bar{3}k|1\right\rangle\bar{u}_{\lambda_q}(k_1)\times\label{qqamp}\\
&&(t^{c_1}_{ij}t^{c_2}_{jk}b^{\mu\nu}(k_1,k_2)-
t^{c_2}_{kj}t^{c_1}_{ji}\bar{b}^{\mu\nu}(k_2,k_1))v_{\lambda_{\bar{q}}}(k_2),
\nonumber
\end{eqnarray}
where $E_{p,cms}=\sqrt{s}/2$ is the c.m.s. proton energy, $t^c$ are
the color group generators in the fundamental representation,
$u(k_1)$ and $v(k_2)$ are on-shell quark and antiquark spinors,
respectively, $b,\,\bar{b}$ are the effective vertices arising from
the Feynman rules in quasi-multi-Regge kinematics (QMRK) approach
illustrated in Fig.~\ref{flvertex}:
\begin{eqnarray} \label{bb}
b^{\mu\nu}(k_1,k_2)=\gamma^{\nu}\frac{\hat{q}_{1}-\hat{k}_{1}-m_q}{(q_1-k_1)^2-m^2}
\gamma^{\mu}-\frac{\gamma_{\beta}\Gamma^{\mu\nu\beta}(q_1,q_2)}{(k_1+k_2)^2}
\; , \\
\bar{b}^{\mu\nu}(k_2,k_1)=\gamma^{\mu}\frac{\hat{q}_{1}-\hat{k}_{2}+m_q}{(q_1-k_2)^2-m^2}
\gamma^{\nu}-\frac{\gamma_{\beta}\Gamma^{\mu\nu\beta}(q_1,q_2)}{(k_1+k_2)^2}
\; , \nonumber
\end{eqnarray}
where $\Gamma^{\mu\nu\beta}(q_1,q_2)$ is the effective three-gluon
vertex. These effective vertices were initially proposed for
massless quarks in Ref.~\cite{FL96} and then extended for massive
case in Ref.~\cite{ktfac,HKSST-qq}. The effective $ggg$-vertices are
canceled out when projecting the $q\bar{q}$ production amplitude
Eq.~(\ref{qqamp}) onto the color singlet state, so only the first
two diagrams in Fig.~\ref{flvertex} contribute to the final result
for the production amplitude. Since we will adopt the definition of
gluon polarization vectors proportional to transverse momenta
$q_{1/2\perp}$, i.e. $\varepsilon_{1,2}\sim q_{1/2\perp}/x_{1,2}$
(see below), then we must take into account the longitudinal momenta
in the numerators of effective vertices (see Eq.~(\ref{bb})).
%-------------------------------------------------------------------
%-------------------------------------------------------------------

The SU(3) Clebsch-Gordan coefficient $\left\langle
3i,\bar{3}k|1\right\rangle=\delta^{ik}/\sqrt{N_c}$ in
Eq.~(\ref{qqamp}) projects out the color quantum numbers of the
$q\bar{q}$ pair onto the color singlet state. Factor $1/\sqrt{N_c}$
provides the averaging of the matrix element squared over
intermediate color states of quarks.

Therefore, we have the following amplitude
\begin{eqnarray}\nonumber
&&{}V_{\lambda_q\lambda_{\bar{q}}}^{c_1c_2,\,\mu\nu}=-\frac{g_s^2}{2}\,
\delta^{c_1c_2}\,\bar{u}_{\lambda_q}(k_1)
\biggl(\gamma^{\nu}\frac{\hat{q}_{1}-\hat{k}_{1}-m}
{(q_1-k_1)^2-m^2}\gamma^{\mu}\\&&\qquad\qquad-\gamma^{\mu}\frac{\hat{q}_{1}-
\hat{k}_{2}+m}{(q_1-k_2)^2-m^2}\gamma^{\nu}\biggr)v_{\lambda_{\bar{q}}}(k_2).
 \label{vector_tensor}
\end{eqnarray}
Taking into account momentum conservation and using the gauge
invariance property, we get the following projection to the light
cone vectors (so called ``Gribov's trick'')
\begin{eqnarray*}
V_{\lambda_q\lambda_{\bar{q}}}^{c_1c_2}&=&
n^+_{\mu}n^-_{\nu}V_{\lambda_q\lambda_{\bar{q}},\,\mu\nu}^{c_1c_2}=
\frac{4}{s}\frac{q^{\nu}_1-q^{\nu}_{1\perp}}{x_1}\frac{q^{\mu}_2-q^{\mu}_{2\perp}}{x_2}
V^{c_1c_2}_{\lambda_q\lambda_{\bar{q}},\,\mu\nu}\\&&=
\frac{4}{s}\frac{q^{\nu}_{1\perp}}{x_1}\frac{q^{\mu}_{2\perp}}{x_2}
V^{c_1c_2}_{\lambda_q\lambda_{\bar{q}},\,\mu\nu}.
\end{eqnarray*}
Last expression shows that an important consequence of the gauge
invariance is the vanishing of the matrix element of the effective
$ggq\bar{q}$-vertex between on-mass-shell quark and antiquark states
in the limit of small $q_{1\perp}$ and $q_{2\perp}$
\cite{ktfac,HKSST-qq}
\begin{eqnarray}\label{GI}
V_{\lambda_q\lambda_{\bar{q}}}^{c_1c_2}\to0\quad\mathrm{for}\quad
q_{1\perp}\;\mathrm{or}\;q_{2\perp}\to0,
\end{eqnarray}

The amplitude (\ref{vector_tensor}), projected out onto a particular
quarkonium state, was successfully applied for the description of
the recent CDF data on the exclusive production of charmonia in
Ref.~\cite{PST_chic}. Here we apply the same formalism for separate
$b$ and $\bar b$ jets production. This is exactly the same formalism
as used recently for the exclusive open charm production in
Ref.~\cite{MPS10}.

In the experimentally important case of the forward proton
scattering ${\bf p}'_{1,2t}\to 0$ we have ${\bf q}_{1,t}\simeq -{\bf
q}_{2,t} \simeq {\bf q}_{0,t}\equiv {\bf q}_t$, and the transverse
mass of the $b{\bar b}$ pair in terms of $b$-quark rapidities
$y_{b,\bar b}$ reads
\begin{eqnarray}\label{Mqq}
M_{b\bar{b},t}^2=M_{b\bar{b}}^2=2m_{b,t}^2(1+\cosh(y_b-y_{\bar b}))
\end{eqnarray}
If one looks at centrally produced $b$-jets only, i.e. $y_{b,\bar
b}\to 0$, then according to Eq.~(\ref{Mqq}) the only way to produce
the large invariant mass $M_{b\bar{b}}\sim M_H$ is to consider
high-$k_t$ jets limit $k_t\gg m_b,q_t$. The amplitude $V_{+-}$
vanishes for jets with very small rapidities, i.e. when $y_b\sim
y_{\bar b}\to 0$. Amplitude $V_{++}$ behaves as $\sim
q_t^2\cos\phi/k_t^2$ and thus extremely suppressed in the high-$k_t$
limit. This is in agreement with the statement that Higgs CEP
background is suppressed in very forward and quark massless limits
for centrally produced $b{\bar b}$ jets (with small rapidities), and
agrees with the $J_z=0$ selection rule \cite{Khoze:2000jm}.

However, the particular high-$k_t$ limit is only a part of the whole
story. In our previous analysis of the exclusive open charm
production in Ref.~\cite{MPS10} it was shown that the dominant
contribution to the $c{\bar c}$ dijet cross section comes from
relatively small quark transverse momenta $k_t\simeq 1$ GeV. The
same should hold for exclusive $b{\bar b}$ pair production relevant
for the Higgs background. Indeed, resolving relation (\ref{Mqq})
with respect to typical rapidity difference $|y_b-y_{\bar b}|=\Delta
y$ neglecting quark transverse momenta $k_t$ and keeping only the
$b$-mass contributions $m_b\simeq 4.5$ GeV at fixed
$M_{b\bar{b}}=120$ GeV, we get $\Delta_y\simeq 6.6$. So, the
irreducible background for Higgs CEP can be dominated by $b$-quarks
with comparably small transverse momenta $k_t\ll m_b$, but with
rather large rapidities $y_b\sim -y_{\bar b} \simeq 3.3$.

In the last kinematical situation the amplitudes
$V_{\lambda_b\lambda_{\bar{b}}}$ are not suppressed by a large
denominators, and significant contributions can be obtained. For
simplicity, keeping only the quark mass $m_b\gg k_t$ and large
rapidity $y_b\sim -y_{\bar b}\sim 3$ contributions, we get for
helicity amplitudes
\begin{eqnarray}\nonumber
&&V_{++}\simeq\frac{ig_s^2|\rho|}{m_b^3\sqrt{\cosh(y_b)\cosh(y_{\bar
b})}} \Big[|\rho|\kappa-2\rho|\kappa|\cos\phi\Big],\\ &&V_{+-}\simeq
\frac{-g_s^2|\rho|^2\tanh(y_b-y_{\bar
b})}{m_b^2\sqrt{\cosh(y_b)\cosh(y_{\bar b})}} \label{lowpt}
\end{eqnarray}
where we have introduced the shorthand complex notations in the
forward limit
\begin{eqnarray}\nonumber
\kappa=k^y+ik^x,\quad \rho=q^y+iq^x,
\end{eqnarray}
for the quark and gluon transverse momenta, respectively.

In this low-$k_t$ regime $V_{+-}$ helicity amplitude is dominated
over $V_{++}$ as opposite to the high-$k_t$ case. From
Eq.~(\ref{lowpt}) we see that the amplitude is proportional to
$q_t^2=-|\rho|^2$, which typically can be of the order of few GeV at
LHC energy. This means that numerically low-$k_t$ contribution (of
course, at not extremely large $y_{b,{\bar b}}$) in the case of
$b$-jets can lead to a dominant contribution to the exclusive
background for Higgs CEP. In this asymptotics, the quark mass $m_b$
plays an important role since it comes into the denominator in
Eq.~(\ref{lowpt}). Precise evaluation of the corresponding signal,
however, demands employing the formulae for the hard amplitudes in
the general form in Eq.~(\ref{vector_tensor}). More detailed
analytical and numerical investigation of contributions from
different parts of the phase space will be presented elsewhere
\cite{MPS_Higgs_large}.

In parallel to the total cross section, we calculate the
differential cross sections for exclusive Higgs boson production.
Compared to the standard KMR approach here we calculate the
amplitude with the hard subprocess $g^*g^*\to H$ taking into account
off-shellness of the active gluons, i.e. fully consistent with the
exclusive production of the $b\bar b$ pairs, where the gluon
transverse momenta play crucial role. The details of the off-shell
matrix element can be found in Ref.~\cite{PTS_ggH_vertex}. In
contrast to the exclusive $\chi_c$ production \cite{PST_chic}, due
to a large factorization scale $\sim M_H$ the off-shell effects for
$g^*g^*\to H$ give only a few percents to the final result.

The same unintegrated gluon distributions based on the collinear
distributions are used for the Higgs and continuum $b{\bar b}$
production. This is absolutely necessary for proper estimate of the
signal-to-background ratio, the main purpose of the present Letter.
In the case of exclusive Higgs production we calculate the
four-dimensional distribution in the standard kinematical variables:
$y, t_1, t_2$ and $\phi$. Assuming for this presentation the full
coverage for outgoing protons\footnote{The exact comparison with
experimental set up requires inclusion of extra cuts on fractional
momentum loss of protons and is slightly different for ATLAS and
CMS.} we construct the two-dimensional distributions $d \sigma / dy
d^2 p_t$ in Higgs rapidity and transverse momentum. The distribution
is used then in a simple Monte Carlo code which includes the Higgs
boson decay into the $b {\bar b}$ channel. It is checked
subsequently whether the $b$ and $\bar b$ enter into the
pseudorapidity region spanned by the central detector. Including the
simple cuts we construct several differential distributions in
different kinematical variables.

%-----------------------------------------------------------------
\begin{figure}
\includegraphics[width=8cm]{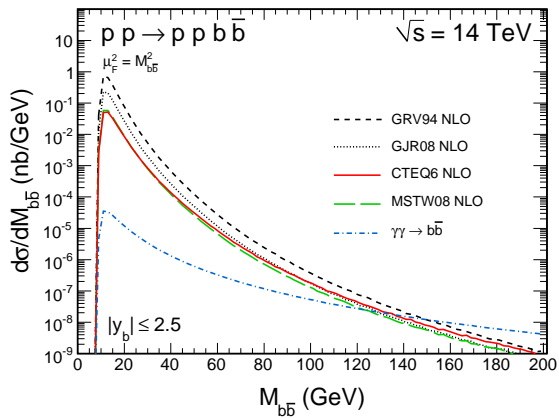}
\caption{ The $b \bar b$ invariant mass distribution for $\sqrt{s}$
= 14 TeV and for $-2.5 < y < 2.5$ corresponding to the ATLAS/CMS
detectors. The absorption effects were taken into account by
multiplying by the gap survival factor $S_G$ = 0.03.}
\label{fig:dsigma_dMbb_PDFs}
\end{figure}
%-----------------------------------------------------------------

\begin{figure}[h!]
 \centerline{
\includegraphics[width=0.33\textwidth]{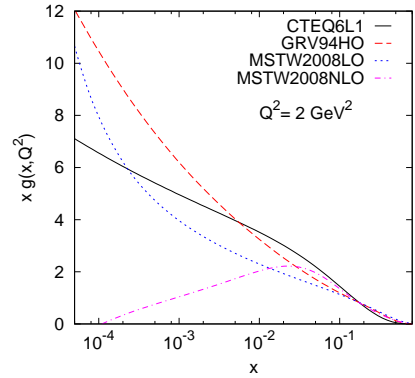}
}
   \caption{\label{fig:pdfs}
   Illustration of the gluon densities as functions of the
   longitudinal momentum fraction $x$
   at the characteristic factorisation scale $Q^2=2$ GeV$^2$ (typical
   for the central exclusive production processes) given by
   the global parameterizations
   CTEQ6L1~\cite{CTEQ}, GRV94HO~\cite{GRV94},
   MSTW2008LO and NLO~\cite{MSTW}.}
\end{figure}

In general, employing the diagonal UGDF in the form (\ref{ugdfkmr})
one encounters a problem of poorly known gluon PDFs at rather low
$x_{1,2}$ and especially small gluon virtualities $q_{\perp}^2$. For
an illustration of the corresponding uncertainties, in
Fig.~\ref{fig:pdfs} we show several parameterizations for the gluon
PDFs widely used in the literature as functions of momentum fraction $x$
at the evolution scale $\sim q_{\perp}^2$ fixed at characteristic value
2 GeV$^2$ typical for the exclusive production of Higgs boson (for
quarkonia production it is even smaller leading to huge
uncertainties as discussed e.g. in Ref.~\cite{PST_chic}). We see
that at $x\lesssim 10^{-3}$ the PDF uncertainties may strongly
affect predictions for not sufficiently large gluon transverse
momenta. In this sense, precise data on the diffractive and
central exclusive production can be used to constrain
the PDF parameterizations \cite{PST_chic,PEI10}.

Testing other models of UGDFs different from Eq.~(\ref{ugdfkmr}) may
be important for estimation of an overall theoretical uncertainty of
our predictions and their stability, and it is planned for our
future study.
%======================================
%======================================

In the following we shall present the main results.
A more detailed analysis with the presentation of
several differential distributions will be given elsewhere
\cite{MPS_Higgs_large}. In Fig.~\ref{fig:dsigma_dMbb_PDFs} we show
the most essential distribution in the invariant mass of the centrally
produced $b \bar b$ pair, which is also being the missing mass of
the two outgoing protons. In this calculation we have taken
into account typical detector limitations in rapidity
$-2.5<y_{b},y_{\bar b}<2.5$.
%and no extra cuts are included.
We show results with different collinear gluon distributions from
the literature: GRV \cite{GRV94}, CTEQ \cite{CTEQ}, GJR \cite{GJR}
and MSTW \cite{MSTW}. The results obtained with radiatively
generated gluon distributions (GRV, GJR) allow to use low values of
$Q_t = q_{0,t}, q_{1,t}, q_{2,t}$ whereas for other gluon
distributions an upper cut on $Q_t$ is necessary. The integrated
double-diffractive $b \bar b$ contribution calculated here seems
bigger than the contribution of the exclusive photoproduction of $b
\bar b$ estimated in \cite{GM07} and details require systematic
studies in the future. The lowest curve in
Fig.~\ref{fig:dsigma_dMbb_PDFs} represents the $\gamma \gamma$
contribution (the bottom diagram in Fig.~\ref{fig:diagrams}). While
the integrated over phase space $\gamma \gamma$ contribution is
rather small, is significant compared to the double-diffractive
component at large $M_{b \bar b}>$ 100 GeV. This can be understood
by damping of the double diffractive component at large $M_{b \bar
b}$ by the Sudakov form factor \cite{KMR_Higgs,MPS10}. In addition,
in contrast to the double-diffractive component the absorption for
the $\gamma\gamma$ component is very small and in practice can be
neglected.

%-----------------------------------------------------------------
\begin{figure}
\includegraphics[width=7.0cm]{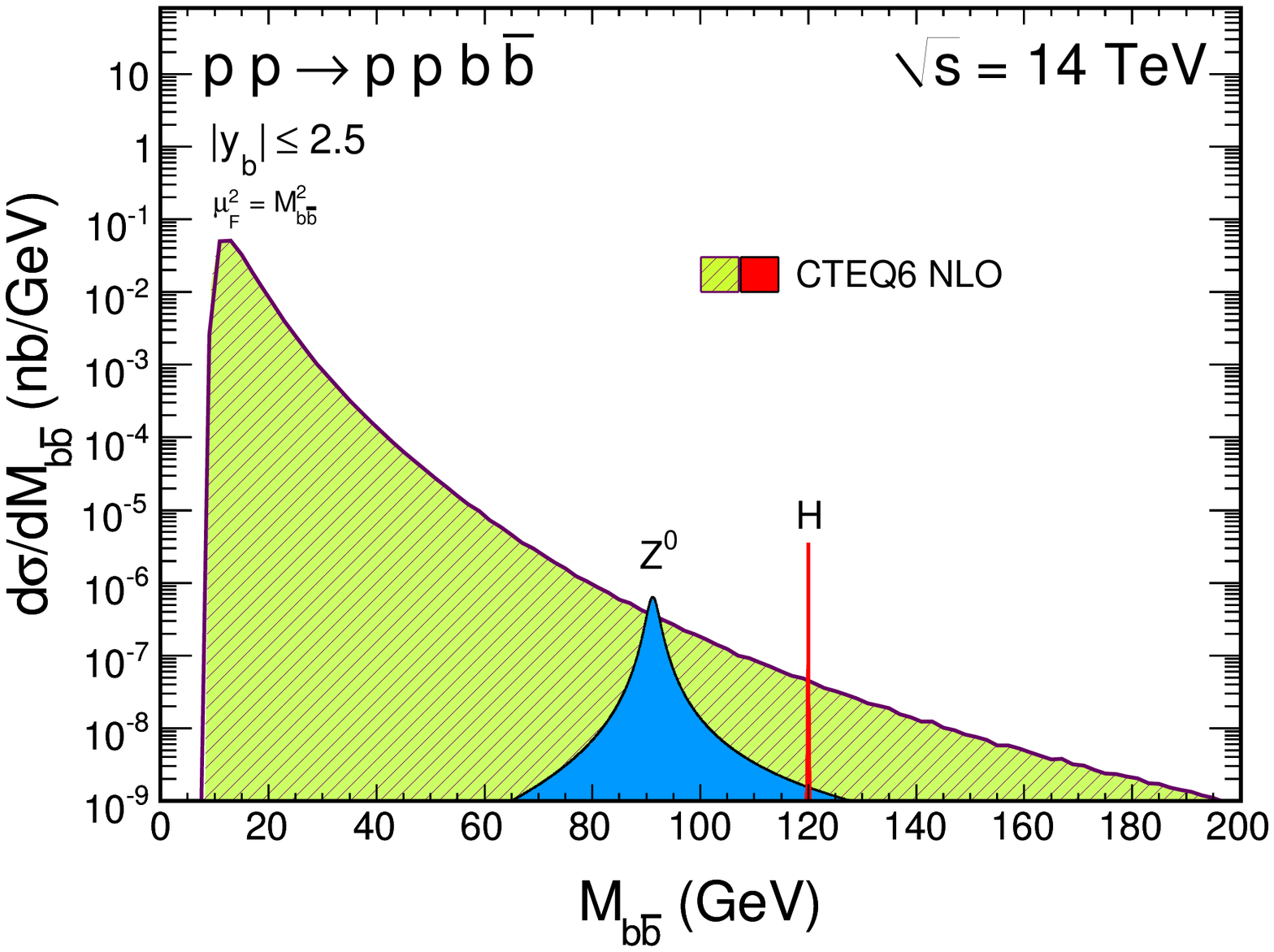}
\includegraphics[width=7.0cm]{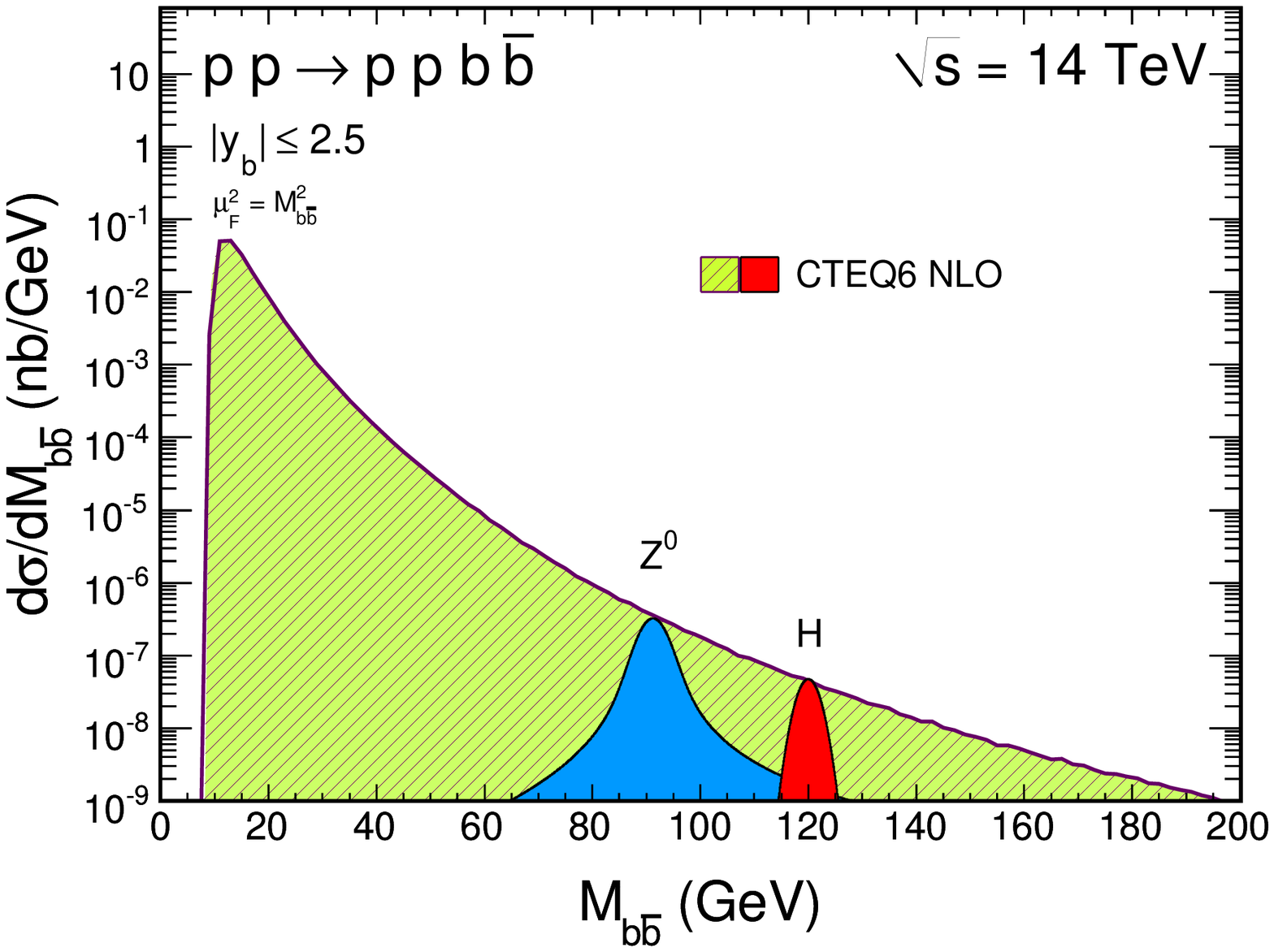}
\caption{The $b \bar b$ invariant mass distribution for $\sqrt{s}$ =
14 TeV and for $b$ and $\bar b$ jets in the rapidity interval $-2.5
< y < 2.5$ corresponding to the ATLAS detector. The absorption
effects for the Higgs boson and the background were taken into
account by multiplying by the gap survival factor $S_G$ = 0.03. The
top panel shows purely theoretical predictions, while the bottom
panel includes experimental effects due to experimental uncertainty
in invariant mass measurement.}
%The left peaks (bumps) correspond to
%the $Z^0$ contribution and the right ones to the Higgs contribution.}
\label{fig:dsigma_dMbb_fully}
\end{figure}
%---------------------------------------------------
In the top panel of Fig.~\ref{fig:dsigma_dMbb_fully} we show the
double diffractive contribution for a selected (CTEQ6 \cite{CTEQ})
collinear gluon distribution and the contribution from the decay of
the Higgs boson including natural decay width calculated as in
Ref.~\cite{Passarino_decay_width}, see the sharp peak at $M_{b \bar
b}$ = 120 GeV (assumed arbitrarily for illustration), which is not
excluded at present by the Higgs searches at LEP
\cite{LEP_Higgs_searches} and Tevatron
\cite{Tevatron_Higgs_searches}. The phase space integrated cross
section for the Higgs production, including absorption effects with
$S_G = 0.03$ is slightly less than 1 fb which is similar to that
predicted by the KMR group. This value is similar as in many KMR
evaluations \cite{KMR_Higgs}. The result shown in
Fig.\ref{fig:dsigma_dMbb_fully} includes also the branching fraction
for BR($H \to b \bar b) \approx$ 0.8 and the rapidity restrictions.
The second much broader Breit-Wigner type peak corresponds to the
exclusive production of the $Z^0$ boson with the cross section
calculated as in Ref.~\cite{CSS09}. The exclusive cross section for
$\sqrt{s}$ = 14 TeV is 16.61 fb including absorption (28.71 fb
without absorption effects). The branching fraction BR($Z^0 \to b
\bar b) \approx$ 0.15 has been included in addition. In contrast to
the Higgs case the absorption effects for the $Z^0$ production are
much smaller \cite{CSS09}. The sharp peak corresponding to the Higgs
boson clearly sticks above the background. In the above calculations
we have assumed an ideal no-error measurement.

In reality the situation is, however, much worse as both protons
and in particular $b$ and $\bar b$ jets are measured with a certain
precision which automatically leads to a smearing in $M_{b \bar b}$ .
While such a smearing is negligible for
the background, it leads to a significant modification of the
Breit-Wigner peaks, especially of the sharp one for the Higgs
boson. In the present Letter the experimental effects are included
in the simplest way by a convolution of the theoretical
distributions with the Gaussian smearing function
\begin{equation}
G(M) = \frac{1}{\sqrt{2 \pi} \sigma}
\exp\left( \frac{ (M-M_H)^2 } { 2 \sigma^2 } \right) \; ,
\end{equation}
with $\sigma$ = 2 GeV, which realistically represents
the experimental situation \cite{Pilkington_private,Royon_private}
and is determined mainly by the precision of measuring forward
protons.
In the bottom panel we show the invariant mass distribution when
the invariant mass smearing is included.
Now the bump corresponding to the
Higgs boson is below the $b \bar b$ background. With the
experimental resolution assumed above the identification of the
Standard Model Higgs will be rather difficult. The situation for
some scenarios beyond the Standard Model may be better
\cite{MPS_Higgs_large}.

%------------------------------------------------------------
\begin{figure}
\includegraphics[width=4.0cm]{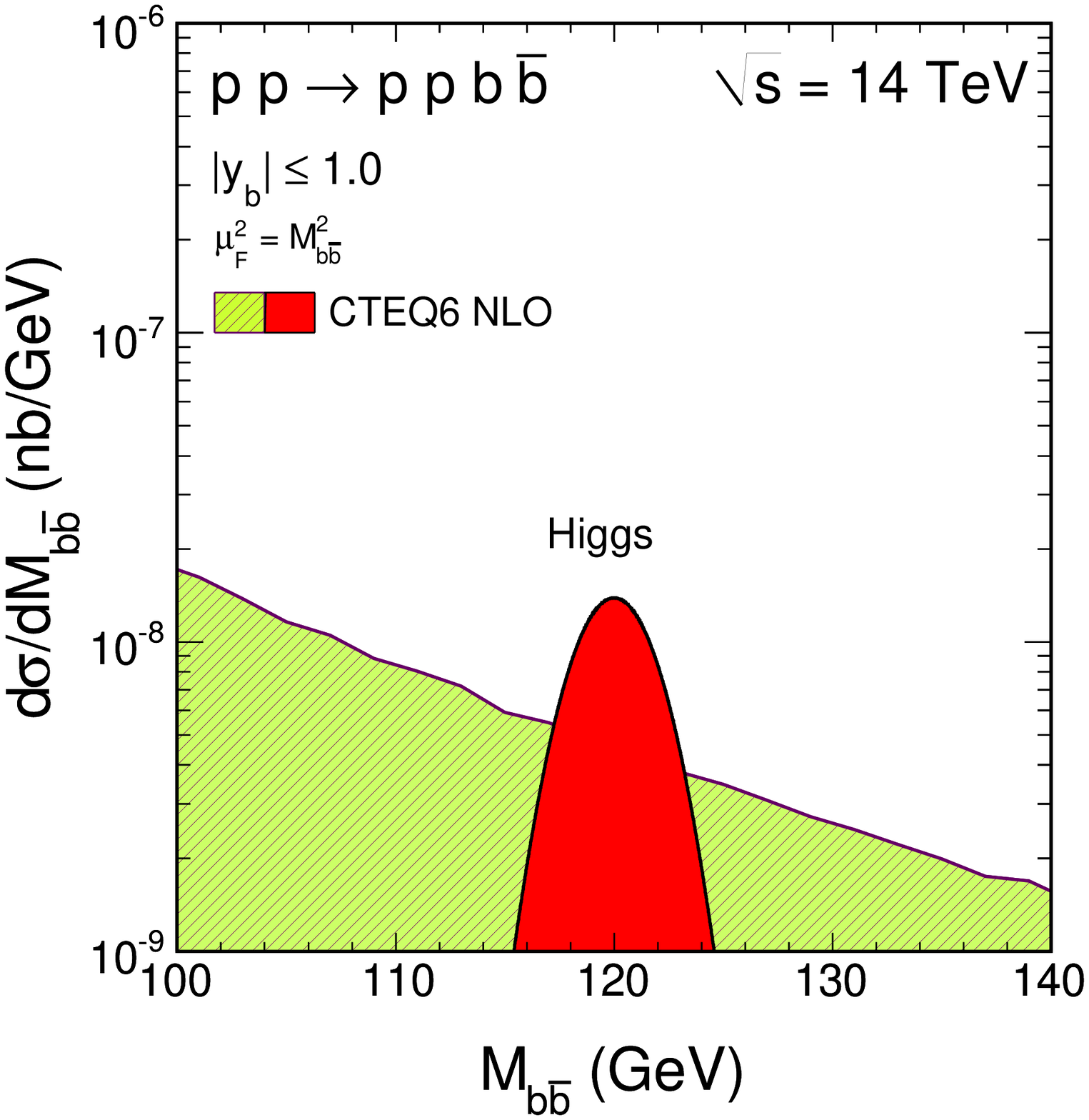}
\includegraphics[width=4.0cm]{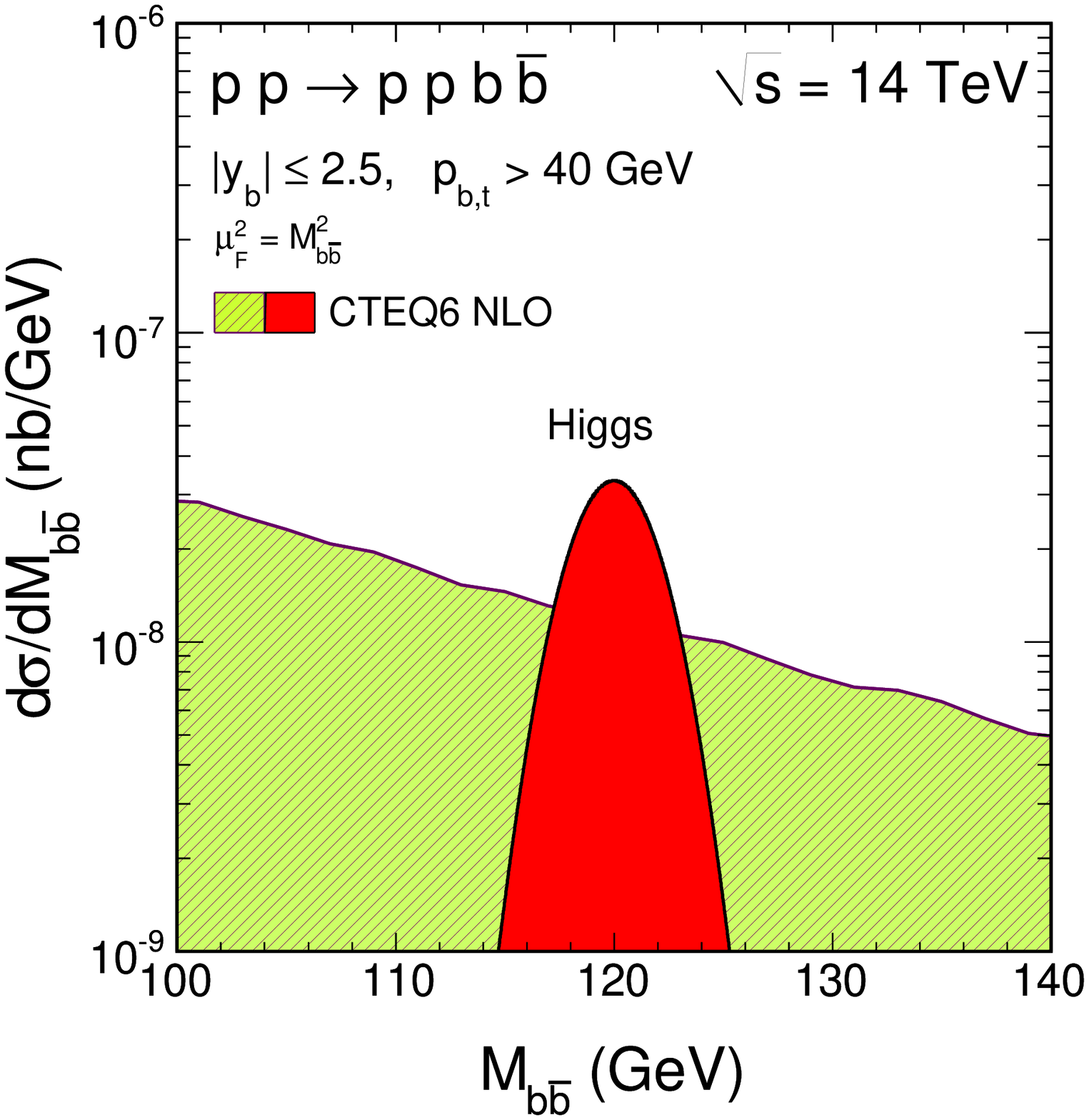}
\caption{The $b \bar b$ invariant mass distribution for $\sqrt{s}$ =
14 TeV for a limited range of $b$ and $\bar b$ rapidities: $-1 <
y < 1$ (left) and for $p_{1,2t} <$ 0.4 GeV (right).}
\label{fig:dsigma_dMbb_cuts}
\end{figure}
%-------------------------------------------------------------
The question now is whether the situation can be improved by
imposing further cuts. In Fig.~\ref{fig:dsigma_dMbb_cuts} (left
panel) we show the result for a more limited range of $b$ and $\bar
b$ rapidity, i.e. not making use of the whole coverage of the main
LHC detectors. Here we omit the $Z^0$ contribution and concentrate
solely on the Higgs signal. Now the signal-to-background ratio is
somewhat improved. This would be obviously at the expense of a
deteriorated statistics. Similar improvements of the
signal-to-background ratio can be obtained by limiting transverse
momenta of outgoing protons (right panel).

%----------------------------------------
\begin{figure}[tbh]
 \centerline{\includegraphics[width=0.27\textwidth]{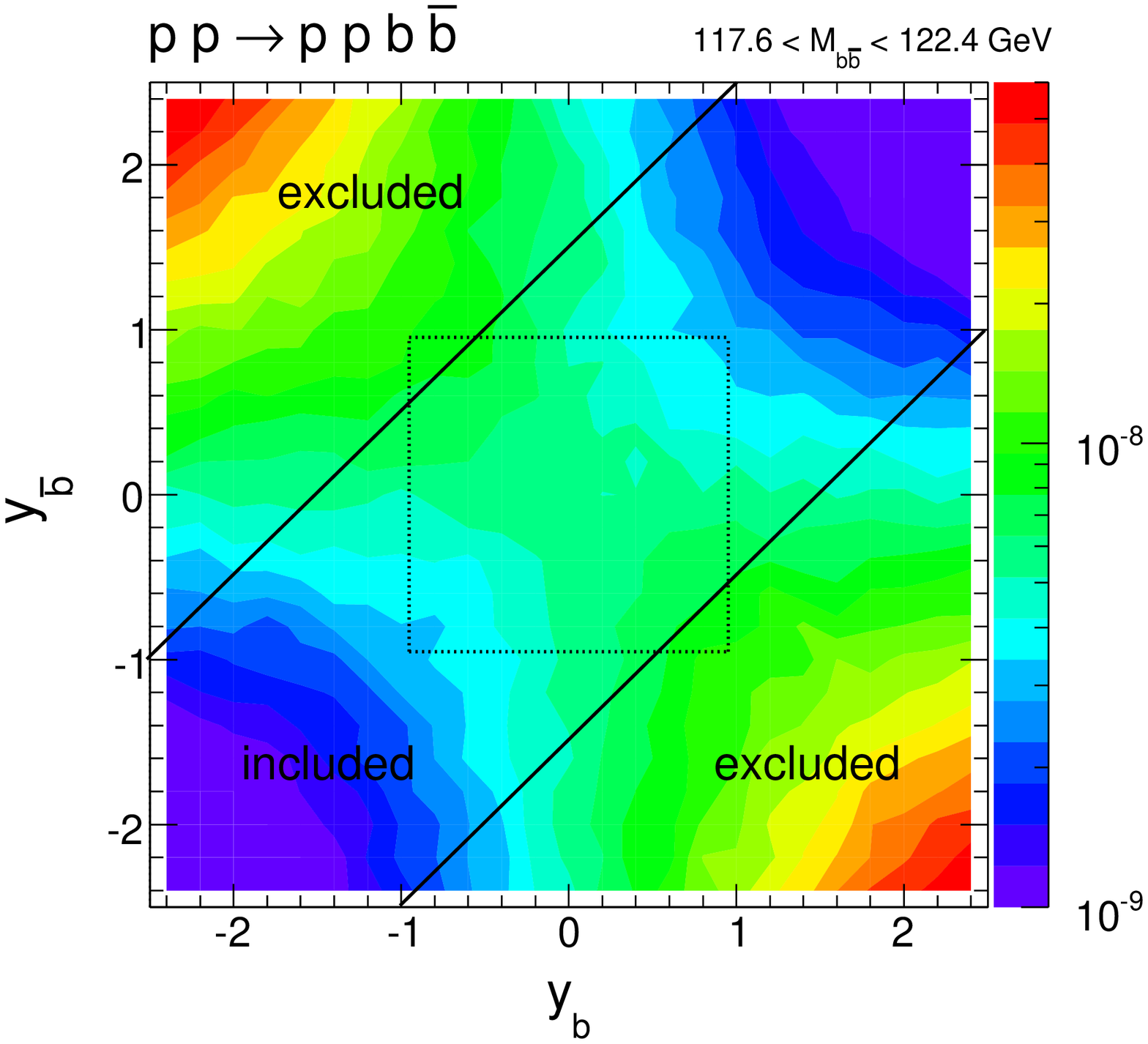}
 \includegraphics[width=0.27\textwidth]{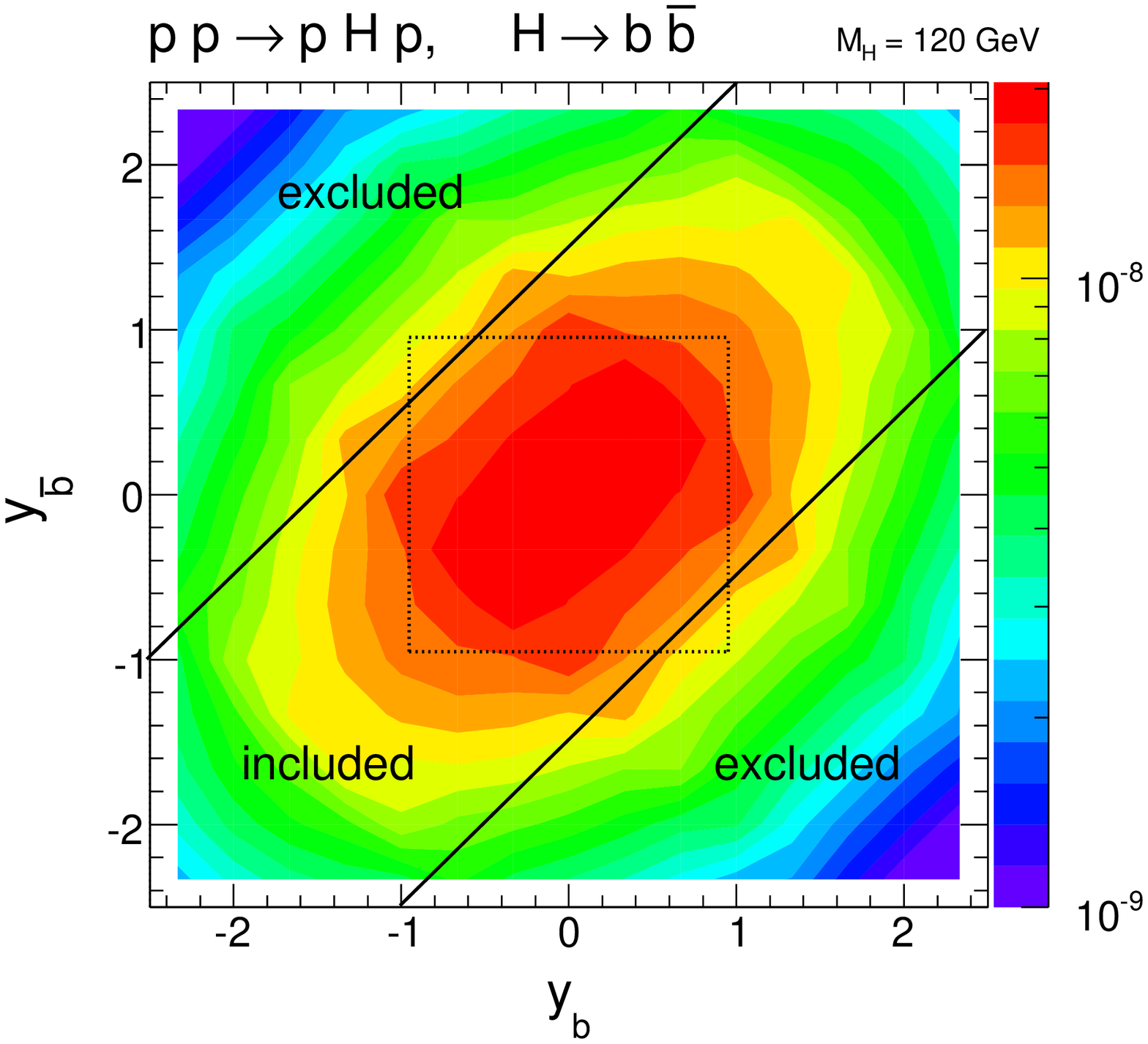}}
   \caption{
 \small
 Two-dimensional distributions over $b$ and ${\bar b}$ rapidities
 for QCD background (left panel), and Higgs CEP signal (right panel)
 with the optimal two-dimensional cut in the $(y_b,\,y_{\bar b})$
 space is marked by the thick contours.}
 \label{fig:cuts-y}
\end{figure}
%---------------------------------------
In order to preserve statistics and to remove most of the $b\bar
b$ background we have to impose more specific two-dimentional cuts.
An example is shown in Fig.~\ref{fig:cuts-y}. Indeed, considering only
the band between two thick solid lines would remove most of the $b \bar b$
background, concentrated mainly in regions with relatively large
difference between quark and antiquark rapidities $|y_b-y_{\bar
b}|\gtrsim1$ (see, the left panel in Fig.~\ref{fig:cuts-y}).
%This contribution is given by low quark $k_t$ asymptotics described by
%Eq.~(\ref{lowpt}).
At the same time, such a cut allows to keep most
of the Higgs signal, which is concentrated in the central rapidity
region $y_{b,\bar{b}} \approx$ 0, as oppose to the background
(see, the right panel in Fig.~\ref{fig:cuts-y}).

In the present analysis we have not been interested in the precise
estimation of the cross section but rather in understanding the
signal-to-background ratio which is of the major importance for the
upcoming Higgs boson searches at the LHC. Consequently, we have
presented results with only one UGDF. This ratio is practically the
same for other UGDFs, which will be shown explicitly in
\cite{MPS_Higgs_large}. The absorption effects have been included
here in a simple multiplicative form. They are expected to be the
same both for the signal and the background, and thus are not
affecting the ratio under consideration. The same gap survival
factor has been used in both cases.

The overall theoretical uncertainty of our predictions for the
absolute value of the $b\bar b$ background contribution is estimated
to be the same as for the Higgs CEP, and given by a factor of 3
more/less \cite{KMR_Higgs}. Theoretical uncertainty in the
signal-to-background ratio under consideration is typically much
smaller, as the main part of the uncertainty coming from the
normalisation of UGDFs ($R_g$ factor) and absorbtive effects (gap
survival $S_G$ factor) is common for both contributions and thus
canceled out in the ratio.

In our analysis we have been concentrated on the irreducible
background only. Other contributions, although, in principle,
reducible, can in practice be also rather troublesome
\cite{Pilkington}. These include dijet misidentification (mainly due
to the $g g \to g g$ subprocess), inclusive double-pomeron processes
\cite{Pomeron-Pomeron-Higgs} and multi-event effects related to
large luminosity \cite{Pilkington}. Further analyses, especially for
the Standard Model Higgs boson production, seem to be necessary to
understand whether the Higgs boson can be identified in the
exclusive production, perhaps not only in the $b \bar b$ decay
channel. The present parton level analysis should be supplemented in
the future by additional analysis of $b \bar b$ jets by including a
model of hadronization. Then standard jet algorithms could be
imposed and the quality of the $b$ and $\bar b$ kinematical
reconstruction could be studied in detail.

\vspace{0.3cm}

%-------------------------------------------------------
We are indebted to Valery Khoze, Oleg Teryaev for a discussion of
central diffraction, to Andy Pilkington
and Christophe Royon for a discussion of experimental limitations of
the exclusive processes discussed in this letter and Ania Cisek for
providing us numerical results of the exclusive $Z^0$ production
from Ref.~\cite{CSS09}. This work was partially supported by the
Polish grant of MNiSW N N202 249235 and by the Carl Trygger
Foundation.
%-------------------------------------------------------

\end{document}